\documentclass[11pt]{article}
\usepackage{amsmath,amsthm,amsfonts}
\usepackage{fullpage}
\usepackage{graphicx}

\newcommand{\C}{\mathbb{C}}
\newcommand{\R}{\mathbb{R}}
\DeclareMathOperator{\tr}{tr}

\begin{document}

\theoremstyle{definition}
\newtheorem{definition}{Definition}
\newtheorem{theorem}{Theorem}
\newtheorem{lemma}{Lemma}
\newtheorem{claim}{Claim}
\newtheorem{proposition}{Proposition}
\newtheorem{exercise}{Exercise}

\newcommand{\identity}{\raisebox{-7pt}{\includegraphics[height=18pt]{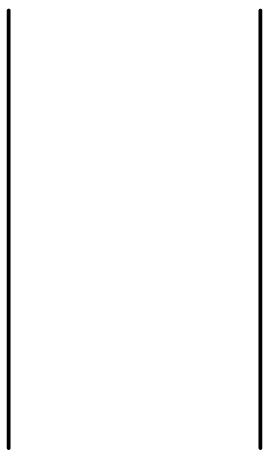}}}
\newcommand{\exchange}{\raisebox{-7pt}{\includegraphics[height=18pt]{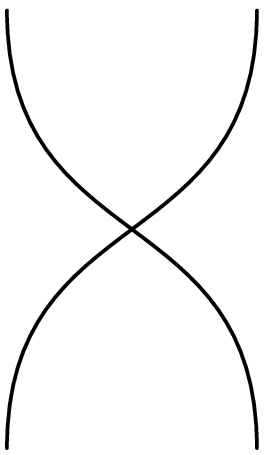}}}
\newcommand{\cupcap}{\raisebox{-7pt}{\includegraphics[height=18pt]{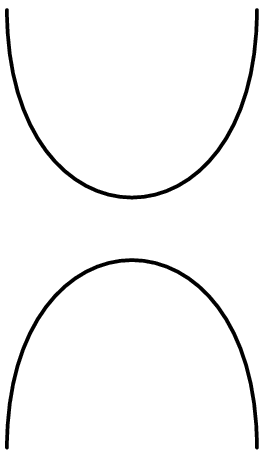}}}

\newcommand{\Z}{{\mathbb{Z}}}
\newcommand{\Exp}{\mathop\mathrm{Exp}}
\newcommand{\Var}{\mathop\mathrm{Var}}
\newcommand{\Cov}{\mathop\mathrm{Cov}}
\newcommand{\abs}[1]{\left| #1 \right|}
\newcommand{\wf}{\widehat{f}}
\newcommand{\wg}{\widehat{g}}
\newcommand{\din}{d^{\rm in}}
\newcommand{\dout}{d^{\rm out}}
\newcommand{\e}{{\rm e}}
\newcommand{\inner}[2]{\left\langle #1 , #2 \right\rangle}
\newcommand{\bra}[1]{\left\langle #1 \right|}
\newcommand{\ket}[1]{\left| #1 \right\rangle}
\newcommand{\norm}[1]{\left| #1 \right|}
\newcommand{\vsym}{V_{\rm sym}}
\newcommand{\pisym}{\Pi_{\rm sym}}
\newcommand{\jun}{j_{\rm undirected}}

\title{Circuit partitions and $\#P$-complete products of inner products}
\author{Cristopher Moore \\ Computer Science Department, \\ University of New Mexico \\ and the Santa Fe Institute
\and
Alexander Russell \\ Department of Computer Science \& Engineering \\
University of Connecticut}
\maketitle

\begin{abstract}
We present a simple, natural $\#P$-complete problem.  Let $G$ be a directed graph, and let $k$ be a positive integer.  We define $q(G;k)$ as follows.  At each vertex $v$, we place a $k$-dimensional complex vector $x_v$.  We take the product, over all edges $(u,v)$, of the inner product $\inner{x_u}{x_v}$.  Finally, $q(G;k)$ is the expectation of this product, where the $x_v$ are chosen uniformly and independently from all vectors of norm $1$ (or, alternately, from the Gaussian distribution).  We show that $q(G;k)$ is proportional to $G$'s cycle partition polynomial, and therefore that it is $\#P$-complete for any $k>1$.
\end{abstract}

\section{Introduction}

\begin{figure}
\begin{center}
\includegraphics[width=1.8in]{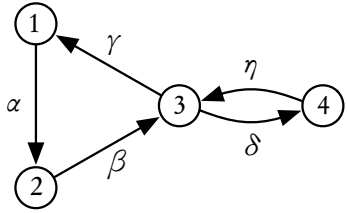}
\end{center}
\caption{A little directed graph.  We use the edge labels in the proof of Theorem~\ref{thm:main}.}
\label{fig:example}
\end{figure}

Let $x, y \in \C^k$ be $k$-dimensional complex-valued vectors.  
We denote their inner product as
\[
\inner{x}{y} = \sum_{i=1}^k x_i^* y_i \, .
\]
Now suppose we have a directed graph $G=(V,E)$.  Let us associate a vector $x_v \in \C^k$ with each vertex $v$, and consider the product over all edges $(u,v)$ of the inner products of the corresponding vectors:
\begin{equation}
\label{eq:prod}
\prod_{(u,v) \in E} \inner{x_u}{x_v} \, .
\end{equation}
For instance, for the graph in Figure~\ref{fig:example} this product is
\begin{equation}
\label{eq:prod-example}
\prod_{(u,v) \in E} \inner{x_u}{x_v} 
= \inner{x_1}{x_2} \inner{x_2}{x_3} \inner{x_3}{x_1} \abs{\inner{x_3}{x_4}}^2 \, . 
\end{equation}
The expectation of this product, where each $x_v$ is chosen independently and uniformly from the set of vectors in $\C^k$ of norm $1$, is a type of moment, where each $x_v$ appears with order $\din_v+\dout_v$.  It is a function of the graph $G$ and the dimension $k$, which we denote as follows:
\[
q(G;k) = \Exp_{\{x_v\}} \prod_{(u,v) \in E} \inner{x_u}{x_v} \, .
\]

A simple observation is that $q(G;k)$ is zero unless $G$ is Eulerian---that is, unless $\din_v = \dout_v$ for each vertex $v$.  Since $x_v$ appears in the product $\din_v$ times unconjugated and $\dout_v$ times conjugated, multiplying $x_v$ by $\e^{i\theta}$ multiplies $q(G;k)$ by $\e^{i\theta (\din_v-\dout_v)}$.  But multiplying by a phase preserves the uniform measure, so the expectation is zero if $\din_v \ne \dout_v$ for any $v$.

So, let us suppose that $G$ is Eulerian.  In that case, what is $q(G;k)$?  Does it have a combinatorial interpretation?  And how difficult is it to calculate?  
Our main result is this:
\begin{theorem}
\label{thm:main}
For any $k \ge 2$, computing $q(G;k)$, given $G$ as input, is $\#P$-hard under Turing reductions.
\end{theorem}
\noindent 
If we extend $\#P$ to rational functions in the natural way, then we can replace $\#P$-hardness in this theorem with $\#P$-completeness.

Our proof is very simple; we show that $q(G;k)$ is essentially identical to an existing graph polynomial, which is known to be $\#P$-hard to compute.  Along the way, we will meet some nice combinatorics, and glancingly employ the representation theory of the unitary and orthogonal groups.

\section{The circuit partition polynomial}

A \emph{circuit partition} of $G$ is a partition of $G$'s edges into circuits.  Let $r_t$ denote the number of circuit partitions containing $t$ circuits; for instance, $r_1$ is the number of Eulerian circuits.  The \emph{circuit partition polynomial} $j(G;z)$ is the generating function
\begin{equation}
\label{eq:j}
j(G;z) = \sum_{t=1}^\infty r_t z^t \, .
\end{equation}
For instance, for the graph in Figure~\ref{fig:example} we have $j(G;z) = z+z^2$.  This polynomial was first studied by Martin~\cite{martin}, with a slightly different parametrization; see also~\cite{arratia,bollobas,bouchet,ellis,jaeger,vergnas79,vergnas88}.

Now consider the following theorem.
\begin{theorem}
\label{thm:q-j}
For any Eulerian directed graph $G=(V,E)$, 
\begin{equation}
\label{eq:q-j}
q(G;k) = 
\left( \prod_{v \in V} \frac{(k-1)!}{(k+d_v-1)!} \right)
j(G;k) \, ,
\end{equation}
where $d_v$ denotes $\din_v = \dout_v$.
\end{theorem}

\begin{proof} 
Given a vector $x \in \C^k$ and an integer $d$, the outer product of $x^{\otimes d} = \underbrace{x \otimes \cdots \otimes x}_{d\ {\rm times}}$ with itself is a tensor of rank $2d$, or equivalently a linear operator on $(\C^k)^{\otimes d}$:
\[
\ket{x^{\otimes d}} \bra{x^{\otimes d}} = \big( \ket{x} \bra{x} \big)^{\otimes d} \, . 
\]
In terms of indices, we can write 
\[
\ket{x^{\otimes d}} \bra{x^{\otimes d}}^{\alpha_1 \alpha_2 \cdots \alpha_d}_{\beta_1 \beta_2 \cdots \beta_d} 
= \prod_{\ell=1}^d x_{\alpha_\ell} x_{\beta_\ell}^* \, . 
\]

Then $\prod_{(u,v) \in E} \inner{x_u}{x_v}$ is a contraction of the product of these tensors, where upper and lower indices correspond to incoming and outgoing edges respectively.  For instance, for the graph in Figure~\ref{fig:example} we can rewrite the product~\ref{eq:prod-example} as
\[
\prod_{(u,v) \in E} \inner{x_u}{x_v}
= \ket{x_1} \bra{x_1}^\gamma_\alpha
\; \ket{x_2} \bra{x_2}^\alpha_\beta
\; \ket{x_3 \otimes x_3} \bra{x_3 \otimes x_3}^{\beta \eta}_{\gamma \delta}
\; \ket{x_4} \bra{x_4}^\delta_\eta \, . 
\]
Here we use the Einstein summation convention, where any index which appears once above and once below is automatically summed from $1$ to $k$.  Now, since the $x_v$ are independent for different $v$, we can compute $q(G;k)$ by taking the expectation over each $x_v$ separately.  This gives a contraction of the tensors
\begin{equation}
\label{eq:xd-def}
X_d = \Exp_{x} \ket{x^{\otimes d}} \bra{x^{\otimes d}} \, ,
\end{equation}
where $d=d_v$, over all $v$.  

In order to calculate $X_d$, we introduce some notation.  Let $S_d$ denote the symmetric group on $d$ elements.  We identify a permutation $\pi \in S_d$ with the linear operator on $(\C^k)^{\otimes d}$ which permutes the $d$ factors in the tensor product.  That is, 
\[
\pi \left( x_1 \otimes x_2 \otimes \cdots \otimes x_d \right) = x_{\pi(1)} \otimes x_{\pi(2)} \otimes \cdots \otimes x_{\pi(d)} \, , 
\]
or, using indices,
\[
\pi^{\alpha_1 \alpha_2 \cdots \alpha_d}_{\beta_1 \beta_2 \cdots \beta_d} 
= \prod_{\ell=1}^d \delta^{\alpha_{\pi(\ell)}}_{\beta_\ell} \, ,  
\]
where $\delta^i_j$ is the Kronecker delta operator, $\delta^i_j=1$ if $i=j$ and $0$ if $i \ne j$.  Diagrammatically, $\pi$ is a gadget with $d$ incoming edges and $d$ outgoing edges, wired to each other according to the permutation $\pi$.

We have the following lemma:
\begin{lemma}
\label{lem:xd}
With $X_d$ defined as in~\eqref{eq:xd-def}, if $x$ is uniform in the set of vectors in $\C^k$ of norm $1$, then
\begin{equation}
\label{eq:xd}
X_d 
= \frac{(k-1)!}{(k+d-1)!} \sum_{\pi \in S_d} \pi \, . 
\end{equation}
\end{lemma}

\begin{proof}
First, $X_d$ is a member of the commutant of the group $U(k)$ of $k \times k$ unitary matrices, since these preserve the uniform measure.  That is $X_d$ commutes with $U^{\otimes d}$ for any $U \in U(k)$.  By Schur duality, the commutant is a quotient of the group algebra $\C[S_d]$; namely, the image of $\C[S_d]$ under the identification above.  Thus $X_d$ is a superposition of permutations, $\sum_{\pi \in S_d} a_\pi \pi$.  

We also have $X_d \pi = \pi X_d = X_d$ for any $\pi$.  Thus $X_d$ is proportional to the uniform superposition on $S_d$, or equivalently the projection operator $\pisym = (1/d!) \sum_\pi \pi$ onto the totally symmetric subspace $\vsym$ of $(\C^k)^{\otimes d}$.  Since $\tr X_d = \Exp_x \abs{x}^{2d} = 1$ while $\tr \pisym = \dim \vsym$, we have $X_d = (1/\dim \vsym) \,\pisym$.  

Finally, $\dim \vsym$ is the number of ways to label the $d$ factors of the tensor product with basis vectors $\{e_1,\ldots,e_k\}$ in nondecreasing order---or, for aficionados, the number of semistandard tableaux with one row of length $d$ and content ranging from $1$ to $k$.  This gives $\dim \vsym = {k+d-1 \choose d}$.

To illustrate some ideas that will recur in the next section, we give an alternate proof.  First, note that $\tr \pi$ is the number of ways to label each of $\pi$'s cycles with a basis vector ranging from $1$ to $k$, or $k^{c(\pi)}$ where $c(\pi)$ denotes the number of cycles (including fixed points).  Thus
\begin{equation}
\label{eq:sd-genfunc}
\tr \sum_{\pi \in S_d} \pi
= \sum_{\pi \in S_d} k^{c(\pi)} \, . 
\end{equation}
To compute this generating function, we use the fact that each permutation $\pi \in S_d$ appears once in the following product, where $1$ denotes the identity permutation, and $\tau_{ij}$ denotes the transposition of the $i$th and $j$th object:
\begin{equation}
\label{eq:sd-sum}
\sum_{\pi \in S_d} \pi = 1 (1 + \tau_{1,2}) (1+\tau_{1,3}+\tau_{2,3}) \cdots (1+\tau_{1,d}+\tau_{2,d}+\cdots+\tau_{d-1,d}) \, . 
\end{equation}
This product works by describing a permutation $\pi_t$ of $t$ objects inductively as a permutation $\pi_{t-1}$ of the first $t-1$ objects, composed either with the identity or with a transposition swapping the $t$th object with one of the previous $t-1$.  If we apply the identity, then the $t$th object is a fixed point, and $c(\pi_t) = c(\pi_{t-1})+1$, gaining a factor of $k$ in~\eqref{eq:sd-genfunc}; but if we apply a transposition, then $c(\pi_t)=c(\pi_{t-1})$.   Thus~\eqref{eq:sd-sum} becomes
\[
\sum_{\pi \in S_d} k^{c(\pi)} = k (k+1) (k+2) \cdots (k+d-1) = \frac{(k+d-1)!}{(k-1)!} \, . 
\]
Comparing traces again gives~\eqref{eq:xd}.
\end{proof}

All that remains is to interpret the operators $X_{d_v}$, and their contraction, diagrammatically.  Lemma~\ref{lem:xd} tells us that, for each vertex $v$ of $G$, taking the expectation over $x_v$ converts it to a sum over all $d_v!$ ways to wire the incoming edges to the outgoing edges.  But doing this at each vertex gives us a sum over all cycle partitions of $G$.  Contracting these tensors gives the number of ways to label each cycle in a each partition with a basis vector ranging from $1$ to $k$, so each cycle contributes a factor of $k$.  Along with the scaling factor in~\eqref{eq:xd}, this completes the proof.
\end{proof}

Next we show that the cycle partition polynomial is $\#P$-hard.  To our knowledge, the following theorem first appeared in~\cite{ellis-sarmiento}; we prove it here for completeness.
\begin{theorem}
\label{thm:nump}
For any fixed $z > 1$, computing $j(G;z)$ from $G$ is $\#P$-hard under Turing reductions.
\end{theorem}

\begin{proof}
Recall that the \emph{Tutte polynomial} of an undirected graph $G=(V,E)$ can be written as a sum over all subsets $S$ of $E$, 
\begin{equation}
\label{eq:tutte}
T(G;x,y) = \sum_{S \subseteq E} (x-1)^{c(S)-c(G)} \,(y-1)^{c(S)+\abs{S}-n} \, . 
\end{equation}
Here $c(G)$ denotes the number of connected components in $G$.  Similarly, $c(S)$ denotes the number of connected components in the spanning subgraph $(V,S)$, including isolated vertices.  When $x=y$, we have
\begin{equation}
\label{eq:tutte2}
T(G;x,x) = \sum_{S \subseteq E} (x-1)^{c(S)+\ell(S)-c(G)} \, ,
\end{equation}
where $\ell(S) = c(s)+\abs{S}-n$ is the total excess of the components of $S$, i.e., the number of edges that would have to be removed to make each one a tree.

\begin{figure}
\begin{center}
\includegraphics[width=3in]{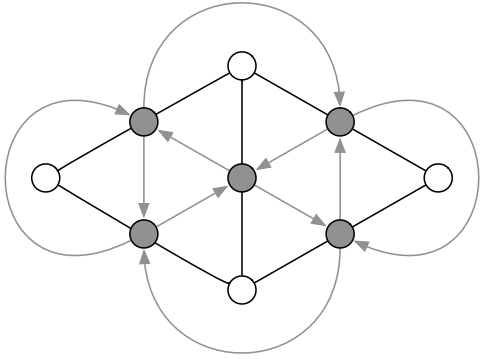}
\end{center}
\caption{A planar graph $G$ (black) and its oriented medial graph $G_m$ (gray).} 
\label{fig:medial}
\end{figure}

If $G$ is planar, then we can define a directed medial graph $G_m$ as in Figure~\ref{fig:medial}.  Each vertex of $G_m$ corresponds to an edge of $G$, edges of $G_m$ correspond to shared vertices in $G$, and we orient the edges of $G_m$ so that they go counterclockwise around the faces of $G$.  Each vertex of $G_m$ has $\din = \dout = 2$, so $G_m$ is Eulerian.

The following identity is due to Martin~\cite{martin}; see also~\cite{vergnas79}, or~\cite{austin} for a review.
\begin{equation}
\label{eq:martin}
j(G_m;z) = z^{c(G)} \,T(G;z+1,z+1) \, . 
\end{equation}
We prove this using a one-to-one correspondence between subsets $S \subseteq E$ and circuit partitions of $G_m$.  Let $v$ be a vertex of $G_m$, corresponding to an edge $e$ of $G$.  Then the circuit partition connects each of $v$'s incoming edge to the outgoing edge on the same side of $e$ if $e \in S$, and crosses over to the other side if $e \notin S$.  It is easy to prove by induction that the number of circuits is then $c(S)+\ell(S)$, in which case~\eqref{eq:tutte2} yields~\eqref{eq:martin}.

The theorem then follows from the fact, proven by Vertigan~\cite{vertigan}, that the Tutte polynomial for planar graphs is $\#P$-hard under Turing reductions, except on the hyperbolas $(x-1)(y-1) \in \{1, 2\}$ or when $(x,y) \in \{(1,1), (-1,-1), (\omega,\omega^*), (\omega^*,\omega)\}$ where $\omega = \e^{2\pi i /3}$.  Thus computing $j(G;z)$ for any $z > 1$ is $\#P$-hard, even in the special case where $G$ is planar and where every vertex has $\din=\dout=2$.
\end{proof}


\section{Real-valued vectors}

We can also consider the case where the $x_v$ are real-valued, and are chosen uniformly from the set of vectors in $\R^k$ of norm $1$.  In this case, the inner product $\inner{x_u}{x_v}$ becomes symmetric, so the graph $G$ becomes undirected.  We might then expect $q(G;k)$ to be related to the circuit partition polynomial for undirected circuits, and indeed this is the case.

We again wish to compute the tensor $X_d = \Exp_x \ket{x^{\otimes d}} \bra{x^{\otimes d}}$.  First, let $M_d$ denote the set of perfect matchings of $2d$ objects; note that 
\[
\abs{M_d} = (2d-1)!! = (2d-1)(2d-3) \cdots 5 \cdot 3 \cdot 1 = \frac{(2d)!}{2^d d!} \, .
\]
We can identify each matching $\mu \in M_d$ with a linear operator on $(\R^k)^{\otimes d}$, where the first $d$ objects correspond to upper indices, and the last $d$ correspond to lower indices.  However, in addition to permutations that wire upper indices to lower ones with a bipartite matching, we now also have ``cups'' and ``caps'' that wire two upper indices, or two lower indices, to each other.  For instance, if $d=2$ then $M_d$ includes three operators, corresponding to the three perfect matchings of $4$ objects:
\begin{equation}
\label{eq:m2}
\delta^{\alpha_1}_{\beta_1} \delta^{\alpha_2}_{\beta_2} = \identity \, , \quad
\delta^{\alpha_1}_{\beta_2} \delta^{\alpha_2}_{\beta_1} = \exchange \, , \quad
\mbox{and} \quad
\delta^{\alpha_1,\alpha_2} \delta_{\beta_1,\beta_2} = \cupcap \, .
\end{equation}
The first two of these operators correspond to the identity permutation and the transposition $\tau_{1,2}$ respectively, as in the previous section.  The third one is a \emph{cupcap}; it is the outer product of the vector $\sum_{i=1}^k e_i \otimes e_i$ with itself, where $e_i$ denotes the $i$th basis vector in $\R^k$.  We denote it $\gamma_{1,2}$, and more generally $\gamma_{ij} = \delta^{\alpha_i,\alpha_j} \delta_{\beta_i,\beta_j}$.

Now, in the real-valued case, Lemma~\ref{lem:xd} becomes the following:
\begin{lemma}
\label{lem:xd-real}
If $x$ is uniform in the set of vectors in $\R^k$ of norm $1$, then
\begin{equation}
\label{eq:xd-real}
X_d 
= \frac{1}{k(k+2)(k+4)\cdots(k+2d-2)} \sum_{\mu \in M_d} \mu
= \frac{(k-2)!!}{(k+2d-2)!!} \sum_{\mu \in M_d} \mu \, , 
\end{equation}
where $n!! = n(n-2)(n-4) \cdots 6 \cdot 4 \cdot 2$ if $n$ is even, and $n(n-2)(n-4) \cdots 5 \cdot 3 \cdot 1$ if $n$ is odd.
\end{lemma}

\begin{proof}
Analogous to the complex case, $X_d$ is a member of the commutant of the group $O(k)$ of $k \times k$ orthogonal matrices, since these preserve the uniform measure.  That is, $X_d$ commutes with $O^{\otimes d}$ for any $O \in O(k)$.  The commutant of $O(k)$ is the \emph{Brauer algebra}; namely, the algebra consisting of linear combinations of the operators $\mu \in M_d$.  Thus $X_d$ is of the form $\sum_{\mu \in M_d} a_\mu \mu$.

In addition to being fixed under permutations as in the complex case, $X_d$ is also fixed under partial transposes, which switch some upper indices with some lower ones.  Thus $X_d$ is proportional to the uniform superposition $\sum_{\mu \in M_d} \mu$.  To find the constant of proportionality, we again compare traces.  

As in the case of permutations, the trace of an operator $\mu \in M_d$ is $k^{c(\mu)}$, where $c(\mu)$ is the number of loops in the diagram resulting from joining the upper indices to the lower ones.  For instance, for the operators in~\eqref{eq:m2}, we have 
$\tr 1 = k^2$, $\tr \tau_{1,2} = k$, and $\tr \gamma_{1,2} = k$.  Thus we wish to calculate
\begin{equation}
\label{eq:md-genfunc}
\tr \sum_{\mu \in M_d} \mu
= \sum_{\mu \in M_d} k^{c(\mu)} \, . 
\end{equation}
We can write $\sum_{\mu \in M_d}$ as a product, analogous to~\eqref{eq:sd-sum}:
\[
\sum_{\pi \in S_d} \pi = 1 (1 + \tau_{1,2} + \gamma_{1,2}) (1+\tau_{1,3}+\gamma_{1,3}+\tau_{2,3}+\gamma_{2,3}) 
\cdots (1+\tau_{1,d}+\gamma_{1,d}+\cdots+\tau_{d-1,d}+\gamma_{d-1,d}) \, . 
\]
This product describes a matching $\mu_t$ of $2t$ objects inductively as a matching $\mu_{t-1}$ of the first $2(t-1)$ objects, composed either with the identity, or with a transposition or cupcap connecting the $t$th upper object with the $i$th lower one and the $t$th lower object with the $i$th upper one, or vice versa.  If we apply the identity, then the $t$th upper object is matched to the $t$th lower one, and $c(\mu_t) = c(\mu_{t-1})+1$, gaining a factor of $k$ in~\eqref{eq:md-genfunc}; but if we apply a transposition or cupcap, then $c(\mu_t)=c(\mu_{t-1})$.   Thus~\eqref{eq:md-genfunc} becomes
\[
\sum_{\pi \in S_d} k^{c(\pi)} = k (k+2) (k+4) \cdots (k+2d-2) = \frac{(k+2d-2)!!}{(k-2)!!} \, . 
\]
We again have $\tr X_d = \Exp_x \abs{x}^{2d} = 1$, and comparing traces gives~\eqref{eq:xd-real}.
\end{proof}

As before, $q(G;k)$ is a contraction of the tensors $X_d$.  However, now $G$ is undirected, with no distinction between incoming and outgoing edges, so at each vertex of degree $d_v$ the appropriate tensor is $X_{d_v/2}$.  Applying Lemma~\ref{lem:xd-real} to each $v$ sums over all the ways to match $v$'s edges with each other, and hence sums over all possible partitions of $G$'s edges into undirected cycles.  The trace of the resulting diagram is again the number of ways to label each cycle with a basis vector.  So, if define a polynomial $\jun(G;z)$ as $\sum_{t=1}^\infty r_t z^t$, where $r_t$ is the number of partitions with $t$ cycles, then Theorem~\ref{thm:q-j} becomes
\begin{theorem}
\label{thm:q-j-real}
For any undirected graph $G=(V,E)$ where every vertex has even degree, if we define $q(G;k)$ by selecting the $x_v$ independently and uniformly from the set of vectors in $\R^k$ with norm $1$, then
\begin{equation}
\label{eq:q-j-real}
q(G;k) = 
\left( \prod_{v \in V} \frac{(k-2)!!}{(k+d_v-2)!!} \right)
\jun(G;k) \, .
\end{equation}
\end{theorem}
\noindent
To our knowledge, the computational complexity of $\jun(G;z)$ is open, although it seems likely that it is also $\#P$-hard.

\section{The Gaussian distribution}

Our results above assume that each $x_v$ is chosen uniformly from the set of vectors in $\C^k$ or $\R^k$ of norm $1$.  Another natural measure would be to choose each component of $x_v$ independently from the Gaussian distribution with variance $1/k$, so that $\Exp[\norm{x_v}^2]=1$.  

For vectors in $\C^k$, the probability density of the norm $\norm{x}$ is then
\begin{equation}
\label{eq:pr}
p\left( \norm{x} \right) = \frac{2k^{k+1}}{k!} \norm{x}^{2k-1} \e^{-k \norm{x}^2} \, , 
\end{equation}
Compared to the case where $\norm{x_v}=1$, each $x_v$ contributes scaling factor of $\norm{x_v}^{2d}$ to the product~\eqref{eq:prod}.  The even moments of~\eqref{eq:pr} are
\[
\Exp\left[ \norm{x}^{2d} \right] = \frac{(d+k-1)!}{k^d (k-1)!} \, , 
\]
so in the Gaussian distribution~\eqref{eq:q-j} becomes 
\begin{equation}
\label{eq:q-j-gaussian}
q(G;k) = 
\left( \prod_{v \in V} \frac{1}{k^{d_v}} \right)
j(G;k) 
= \frac{1}{k^m} \,j(G;k) \, ,
\end{equation}
where $m$ denotes the number of edges.

We could also have derived this directly from the Gaussian analog of Lemma~\ref{lem:xd}.  If $x$ is chosen according to the Gaussian distribution on $\C^k$, and we again let $X_d$ denote $\Exp_x \ket{x^{\otimes d}} \bra{x^{\otimes d}}$, then 
\begin{equation}
\label{eq:xd-gaussian}
X_d
= \frac{1}{k^d} \,\sum_{\pi \in S_d} \pi \, . 
\end{equation}

Similarly, in the real-valued case, if we choose each component of $x \in \R^k$ from the Gaussian distribution on $\R$ with variance $1/k$, then~\eqref{eq:q-j-real} becomes
\begin{equation}
\label{eq:q-j-real-gaussian}
q(G;k)  
= \frac{1}{k^m} \,\jun(G;k) \, ,
\end{equation}
since
\begin{equation}
\label{eq:xd-real-gaussian}
X_d 
= \frac{1}{k^d} \,\sum_{\mu \in M_d} \mu \, . 
\end{equation}
Both~\eqref{eq:xd-gaussian} and~\eqref{eq:xd-real-gaussian} are consequences of Wick's Theorem~\cite{isserlis,wick}, that if $x_1,\ldots,x_{2t}$ obey a multivariate Gaussian distribution with mean zero, then
\[
\Exp\left[ \prod_{i=1}^{2t} x_i \right] = \sum_{\mu \in M_t} \prod_{(i,j) \in \mu} \Exp[x_i x_j] \, . 
\]

\paragraph{Acknowledgments.}  We are grateful to Piotr \'Sniady for teaching us the sum~\eqref{eq:sd-sum}, and to Jon Yard for introducing us to the Brauer algebra.  This work was supported by the NSF under grant CCF-0829931, and by the DTO under contract W911NF-04-R-0009.


\begin{thebibliography}{99}

\bibitem{arratia} 
R. Arratia, B. Bollob\'as, and G. Sorkin, 
\newblock The interlace polynomial: A new graph polynomial.
\newblock \emph{Proc. 11th Annual ACM-SIAM Symposium on Discrete Algorithms} 237--245 (2000).

\bibitem{austin}
Andrea Austin, 
\newblock The Circuit Partition Polynomial with Applications and Relation to the {T}utte and Interlace Polynomials.
\newblock \emph{Rose-Hulman Undergraduate Mathematics Journal}, 8(2) (2007).

\bibitem{bollobas}
B\'ela Bollob\'as, 
\newblock Evaluations of the Circuit Partition Polynomial.
\newblock \emph{Journal of Combinatorial Theory, Series B} 85, 261--268 (2002)

\bibitem{bouchet}
Andr\'e Bouchet,
\newblock {T}utte-{M}artin polynomials and orienting vectors of isotropic systems.
\newblock \emph{Graphs Combin.} 7(3) 235--252 (1991).

\bibitem{brauer}
Richard Brauer, 
\newblock On Algebras Which are Connected with the Semisimple Continuous Groups.
\newblock \emph{Annals of Mathematics}, 38(4) 857--872 (1937).

\bibitem{ellis} 
Joanna A. Ellis-Monaghan, 
\newblock New results for the {M}artin polynomial.
\newblock \emph{Journal of Combinatorial Theory, Series B} 74, 326--352 (1998).

\bibitem{ellis-sarmiento}
Joanna A. Ellis-Monaghan and Irasema Sarmiento, 
\newblock Distance hereditary graphs and the interlace polynomial. 
\newblock \emph{Combinatorics, Probability and Computing} 16(6) 947--973 (2007).

\bibitem{isserlis}
L. Isserlis, 
\newblock On a formula for the product-moment coefficient of any order of a normal frequency distribution in any number of variables.
\newblock \emph{Biometrika} 12: 134--139 (1918).

\bibitem{jaeger}
F. Jaeger, 
\newblock On {T}utte polynomials and cycles of plane graphs.
\emph{Journal of Combinatorial Theory, Series B} 44, 127--146 (1988).

\bibitem{martin}
P. Martin, 
\newblock Enum\'erations eul\'eriennes dans les multigraphes et invariants de {T}utte-{G}rothendieck.
\newblock Thesis, Grenoble 1977.

\bibitem{vergnas79}
Michel Las Vergnas, 
\newblock On {E}ulerian partitions of graphs.
\emph{Research Notes in Mathematics} 34, 62--75 (1979).

\bibitem{vergnas88}
Michel Las Vergnas, 
\newblock On the evaluation at $(3, 3)$ of the {T}utte polynomial of a graph.
\emph{Journal of Combinatorial Theory, Series B} 44, 367--372 (1988).

\bibitem{vertigan}
Dirk Vertigan, 
\newblock The Computational Complexity of {T}utte Invariants for Planar Graphs.
\newblock \emph{SIAM J. Comput.}, 35(3) 690--712 (2006).

\bibitem{wenzl}
Hans Wenzl,
\newblock On the Structure of Brauer's Centralizer Algebras.
\newblock \emph{Annals of Mathematics} 128(1) 173--193 (1988).

\bibitem{wick}
Gian-Carlo Wick, 
\newblock The evaluation of the collision matrix. 
\newblock \emph{Physical Review} 80(2): 268--272 (1950).



  
\end{thebibliography}
\end{document}